\documentclass[twocolumn,eqsecnum,aps]{revtex4} 
\usepackage{amsmath}
\usepackage{graphics}
\usepackage{graphicx}
\usepackage{epsfig}
\usepackage{latexsym}
\newcommand{\pslash}{\;/\hspace{-0.6em}p}
\begin{document}
\title{$\chi$SB and Point-Splitting Procedure in Light-Cone Quantized QED in a Magnetic Field}
\author{Daniela Steinbacher}
\affiliation{Institut f\"ur Theoretische Physik III \\ Universit\"at Erlangen-N\"urnberg \\
Staudtstra\ss e 7 \\ 
D-91058 Erlangen, 
Germany }
\date{\today} 
\begin{abstract}
The summation of all rainbow diagrams in light-cone quantized QED in a strong magnetic field leads as in the standard approach to a dynamical electron mass. Further contributions to this summation however can cause problems with light-cone singularities. It is shown that these problems are generally avoided by applying  the point-splitting regularization to every diagram. The possibility of implementing this procedure into the Lagrangian of the theory is discussed.
\end{abstract} 
\maketitle

\section{Introduction}

One of the main advantages of the light-cone formalism (for a review see e.g.\cite{brodsky1,burk1}) is the simplicity of the vacuum structure. It is the origin of most of the simple properties of field theories when quantized on the light-cone.

The triviality of the vacuum structure on the light-cone poses however conceptual problems with regard to non-perturbative phenomena which are attributed to the existence of a nontrivial vacuum. An example is the breakdown of chiral symmetry. In many attempts (see e.g.\cite{lenz3,burklenz,burkEl,burk96,itakura1}) the question of how different phases of a system can be built upon a vacuum which is determined kinematically like the light-cone vacuum  has been addressed. In light-cone quantum field theory vacuum expectation values are in general calculated as expectation
values of Schr\"odinger operators. The kinematic origin of the vacuum leads therefore even in the presence of interactions to trivial
vacuum expectation values. It was stated that vacuum expectation values of Schr\"odinger
operators cannot be considered as meaningful physical quantities and therefore cannot
serve as order parameters \cite{burklenz,lenz3}. Instead it is more reasonable to define order
parameters like the chiral condensate as vacuum expectation values of equal
light-cone time limits of Heisenberg operators point-split in the light-cone time direction. It is thus possible to obtain the correct chiral condensate in light-cone quantization.
In the same way the derivation of a gap equation on the light-cone and
the corresponding dynamical mass should make use of limits of expectation values of Heisenberg operators. 

In this work I derive chiral symmetry breaking on the light-cone. This program is carried out for QED in a magnetic field, following closely a similar calculation in standard coordinates \cite{gusmilga}. It will be shown that the results received in standard coordinates are obtained by a light-cone calculation. The essential step consists in the summation of all rainbow diagrams. Considering further  contributions to the electron self-energy which are not taken into account in the rainbow approximation exhibits problems with specific light-cone singularities. I investigate the possibility to implement a point-splitting procedure to ensure correct results in the light-cone quantized  theory.

  
\section{$\chi$SB in light-cone quantized QED in a constant magnetic field}

Chiral symmetry breaking in massless QED in a constant magnetic field has been established in standard coordinates as a universal phenomenon in 2+1 and 3+1
dimensions. Either by summation of rainbow diagrams (see Fig.\ref{rainbow}) \cite{gusmilga} or via a Schwinger-Dyson
approach, it is possible to obtain a non-perturbative integral equation
for the mass function of the electron \cite{gus,gusmiransky95,lee}.
For realizing qualitatively the breakdown of chiral symmetry it suffices to take into account the lowest Landau level. In the regime of a very strong magnetic field it yields the dominant contributions to the mass operator. This is also valid if the electron mass is generated dynamically.

\subsection{Mass Operator to 1-Loop Order}
\label{1-loop}

Responsible for the chiral symmetry breaking  are the leading logarithmic contributions of the rainbow diagrams. They provide the main contribution to the mass operator. It is instructive to observe how they are obtained at the 1-loop level. 
I use the following notation for 
coordinates and momenta
$$x^{\pm}=\frac{1}{\sqrt{2}}(x^{0}\pm x^{3}),\quad k_{\pm}=\frac{1}{\sqrt{2}}(k_{0}\pm k_{3})$$
and refer to $x^{+},k_{+}$ as light-cone time and light-cone ener\-gy respectively. The magnetic field points in the $ x^{3}$ direction.
The Green function in a magnetic field in light-cone quantization can be obtained in analogy to the calculation in standard coordinates \cite{schwinger,chodos}.
 To 1-loop order the mass operator $M_1$ is computed by taking account of only the lowest Landau level in the Green function. 
After use of the residue theorem, it acquires the form
\begin{align}
\label{nachres}
M_1=\frac{\alpha m_0}{\pi}\int\limits_{0}^{eB}dz\int\limits_0^{1}\frac{dx}{p_{||}^2 \,x^2-x\left(p_{||}^2+m_0^2-z \right)+m_0^2-i\epsilon}\,.
\end{align}
We substituted $\frac{k_-}{p_-}$ by $x$. The integration over $k_+$ leads straightforwardly to an expression obtainable also in standard coordinates. The specific light-cone problems with divergences for small light-cone momenta do not appear here.  

We divide the mass operator into its imaginary and real part.
 For $p_{||}^2>0$ and after the substitution $xp_{||}^2+\frac{xz}{x-1} \rightarrow
 z$, the real part of the mass operator (\ref{nachres}) becomes 
\begin{align}
\label{real}
 \Re \{M_1(p_{||}^2)\}&=\frac{\alpha m_0}{\pi} \int\limits_0^{1}\frac{dx}{x}\int\limits_{xp_{||}^2}^{\frac{xeB}{1-x}-xp_{||}^2}\frac{dz}{z-m_0^2}\quad,
\end{align}
where the principle value of the $z$-integral was inserted. This integral receives the double logarithmic part, from the region where $x \leq \frac{p_{||}^2}{eB}\ll 1$. The upper bound in the $z$-integral is thus approximated by $xeB$.
Approximating analogously the part for $p_{||}^2<0$ and computing the imaginary part, the complete 1-loop expression on the light-cone is
\begin{align}
\label{M1ende}
M_1(p_{||}^2)=\frac{\alpha m_0}{\pi} &\left[
  \ln\left(\frac{|p_{||}^2|}{m_0^2}\right)\ln\left(\frac{eB}{|p_{||}^2|}\right)+\frac{1}{2}\ln^2\left(\frac{eB}{|p_{||}^2|}\right) \right.\nonumber\\
&\left.+i\pi \ln\left(\frac{p_{||}^2}{m_0^2}\right) \Theta(p_{||}^2)
\right]\;.
\end{align}
A comparison between this solution and a numerical evaluation of the principal
value of (\ref{nachres}) justifies the use of the approximations for the region $m_0^2 \leq |p_{||}^2|\leq eB$.

\subsection{Integral Equation on the Light-Cone}

\begin{figure}[b]
 \begin{center}
  \epsfig{file=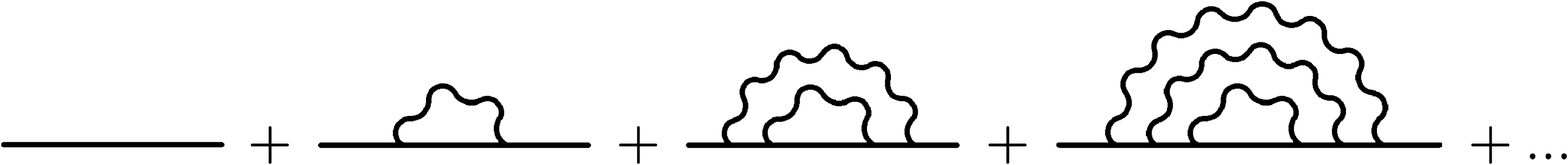, width=8.5cm, height = 1.6cm} 
  \vspace{-0.6cm}
  \caption{Summing up rainbow graphs}
   \label{rainbow}
  \end{center}
 \vspace{-0.4cm}\end{figure}

The integral equation for the mass operator is derived by summation of the leading double-logarithmic contributions coming from the rainbow graphs (see Fig.\ref{rainbow}).
We use a mean field approximation, i.e. we neglect the effects due to the momentum dependence of the mass operator in the denominator of the electron propagator,
\begin{align}
\label{inteq2}
M(p_{||}^2)=m_0&+\frac{\alpha}{i\pi^2} \int\limits_{-\infty}^{\infty} dk_+dk_-
\frac{M(2k_+k_-)}{2k_+k_--m^2+i\epsilon}\cdot \nonumber\\ 
\cdot & \int\limits_0^{eB} \frac{d{\bf k}_{\perp}^2}{2(k_+-p_+)(k_--p_-)-{\bf k}_{\perp}^2+i\epsilon}\,.
\end{align}
We apply a contour integration to do the $k_+$-integral and use the same approximation as in Section \ref{1-loop}. 
The imaginary part of this equation represents processes where real pairs of electrons and photons can be created. In the end we set $|p_{{||} }^{2}=m^2$ where the imaginary part vanishes. It is thus ignored in the following calculation.  
After a partial integration in $z$,
the approximate integral equation becomes
\begin{align}
M(p_{{||} }^{2}) \approx
m_{0}+\frac{\alpha}{2\pi}&\left[\ln\left(\frac{eB}{|p_{{||}
    }^{2}|}\right)\int\limits_{m^{2}}^{|p_{{||} }^{2}|}dx \,\;\frac{M(x)}{x}\right.\nonumber\\ 
&\left.+\int\limits^{eB}_{|p_{{||} }^{2}|}dx \;\frac{M(x)}{x}\;\ln\left(\frac{eB}{x}\right)\right]\quad.
\end{align}
This integral equation resembles the equation in \cite{gusmilga}. 
Its solution leads in the limit $m_0 \to 0$ to a non-zero dynamical mass for an originally massless electron in a constant magnetic field, i.e. to chiral symmetry breaking:
\begin{align}
m=\sqrt{eB}\; e^{-\frac{\pi}{2}\sqrt{\frac{\pi}{2\alpha}}}\quad.
\end{align}
Chiral symmetry breaking can therefore be successfully derived by a summation of all rainbow diagrams in light-cone quantized QED. Like the 1-loop mass operator of Section \ref{1-loop} also higher order rainbow diagrams are not afflicted with light-cone singularities. This yields a derivation of $\chi$SB on the light-cone without additional light-cone problems. 

\section{The Point-Splitting Procedure}
\label{pointsplitting}

Unlike the summation of rainbow diagrams, the evaluation of the exact Schwinger-Dyson equation
\begin{align}
S_B^{-1}(x,y)=G_B^{-1}(x,y)+\Sigma(x,y) \quad,
\end{align}
would require the full electron propagator $S_B$ and therefore the spinor structure of the self-energy $\Sigma$ and contributions apart from rainbow diagrams.
I first discuss the $\gamma$-part of the $1$-loop rainbow
diagram which is proportional to
\begin{align}
\label{gamma-term}
D&=\frac{\alpha}{2\pi^2}\int\limits_0^{eB}dz\;\int d^2k_{||}
\;\frac{\gamma^+k_++\gamma^-k_-}{(k^2_{||}-m_0^2+i\epsilon)[(k-p)^2_{||}-z+i\epsilon]}\,.
\end{align}
A calculation in standard coordinates leads after the insertion of a Feynman para\-meter and a shift of variables to
\begin{align}
\label{standard}
D&=\frac{\alpha}{2\pi}\int\limits_0^{eB}dz\;\int\limits_0^1dx\;\frac{ix\pslash_{||}}{\left[xp_{||}^2(x-1)+xz-(x-1)m_0^2-i\epsilon\right]}\,.
\end{align}
In light-cone coordinates, if one performs first the integration over $k_+$, problems appear, cf.\cite{burkLang} in the  $\gamma^+$ part
\begin{align}
\label{D+ohne}
D_+=\frac{\alpha}{2\pi^2}&\;\int dk_+ dk_-
\;\frac{k_+}{(2k_+k_--m_0^2+i\epsilon)}\cdot \nonumber\\
\cdot&\int\limits_0^{eB}dz\;\frac{1}{[2(k_+-p_+)(k_--p_-)-z+i\epsilon]} \,.
\end{align}
A straightforward evaluation of this integral by a contour integration over $k_+$ yields 
\begin{align}
\label{noncov}
D_+&=\frac{i\alpha}{4\pi p_-}\int\limits_0^{eB}dz\;  \left[\int\limits_{0}^{1}dx \frac{p_{||}^2x}{F^2(x,z)}-\ln{\frac{z}{m_0^2}} \right]\quad,
\end{align}
where $F^2(x,z)=(x-1)xp_{||}^2+xz-(x-1)m_0^2-i\epsilon$ with $x=\frac{k_-}{p_-}$. 
The first integral in (\ref{noncov}) is identical to the $\gamma^+$-part of the covariant result (\ref{standard}). A calculation of the $\gamma^-$-part with $k_+$ integrated first leads on the other hand to the $\gamma^-$-part of the covariant result and no additional term. Therefore the complete expression for the self-energy deviates from the covariant form.  Transforming the integrand in (\ref{D+ohne}) via the algebraic identity \cite{burk1}
\begin{align}
\label{id}
&\frac{2p_-k_+}{(2k_+k_--m_0^2+i\epsilon)[2(k_+-p_+)(k_--p_-)-z+i\epsilon]}= \nonumber\\
&= \frac{2(p_--k_-)p_++m_0^2-z}{(2k_+k_--m_0^2+i\epsilon)[2(k_+-p_+)(k_--p_-)-z+i\epsilon]}\;+\nonumber\\
&+\frac{1}{[2(k_+-p_+)(k_--p_-)-z+i\epsilon]}-\frac{1}{(2k_+k_--m_0^2+i\epsilon)},
\end{align}
the reason for the additive non-covariant part becomes visible. The integrals over the second and third part on the right hand side have the structure of a tadpole diagram in $\varphi^4$-theory. These tadpole integrals are known to cause problems on the light-cone. An application of contour integration leads to a vanishing of the difference of the two integrals. The pole contribution of each integral is zero since the contour can be closed for every $k_-$ such that no pole is enclosed. The surface terms of the integrals cancel each other. This argument however neglects the $k_+$-integral when $k_-=0$ which leads to a $\delta$-function of $k_-$ \cite{chang2}.

After implementing point-splitting exponentials in the tadpole integrals, they remain finite and yield a representation of the modified Bessel function $K_0$,
\begin{align}
\label{BesselK}
\int dk_+ dk_-\frac{e^{ik_+\xi^++ik_-\xi^-}}{2k_+k_--m_0^2+i\epsilon}&=-i\pi\int\limits_0^{\infty}\frac{dk_-}{k_-}\,e^{i\frac{m_0^2}{2k_-}\xi^++ik_-\xi^-} \nonumber\\
&\hspace{-0.0cm}=-2\pi i K_0\left(m_0\xi_L\right)\,,
\end{align}
with $\xi_L=\sqrt{-2\xi^+\xi^-}$. Carrying out the limit $\xi^{\pm} \rightarrow 0$ at the end of the calculation, the two tadpole integrals yield
\begin{align}
D_+^{tad}&=\lim_{\xi^{\pm}\to 0}\frac{-i\alpha}{4\pi
  p_-}\int\limits_0^{eB}
dz\;\left[e^{ip\xi}K_0\left(\sqrt{z}\xi_L\right)-K_0\left(m_0\xi_L\right)\right] \nonumber\\
&=\frac{i\alpha}{4\pi p_-}\int\limits_0^{eB} dz\; \ln{\frac{z}{m_0^2}} \,,
\end{align}
 where the asymptotic behavior of the modified Bessel function for small $z$ has been inserted.
Thus the correctly evaluated tadpole contributions cancel exactly the additional non-covariant part in (\ref{noncov}). The result for the point-split $\gamma^+$-part of the $1$-loop self-energy on the light-cone coincides now with the $\gamma^+$-part of the covariant result (\ref{standard}). 

Obviously the tadpole integrals are not well-defined if the limit $\xi^{+}
\rightarrow 0$ is carried out before the $k_-$-integration
is performed. The dispersion relation on the
light-cone makes a regularization necessary that deals with
the $\frac{1}{k_-}$ singularity for $k_- \rightarrow 0$.
It is therefore the regularization of $k_+$ which is crucial for
obtaining the correct result in light-cone coordinates. It is indispensable to
leave the light-cone time parameter $\xi^+$ finite till the end of the calculation.

The $\delta$-function appearing without point-splitting and which is the source of the whole contribution is now regularized. The contribution from the zero-mode is distributed around $k_-=0$ and the single point $k_-=0$ is no longer important. The point-splitting ensures that we do not have to take care of the zero-mode any longer.

Thus unlike e.g. Pauli-Villars the point-splitting regularization treats the light-cone singularities properly.

\section{Higher Order Diagrams}
\label{HOD}

In this section I show that point-splitting regularizes the light-cone singularities also in higher order diagrams.
The principal structure of $k_{\pm}$-integrals in QED diagrams is determined by the electron and photon propagator. Single parts of these integrals (neglecting overall constants with respect to $k_{\pm}$) can always be written in the form 
\begin{align}
\label{form}
\int dk_- dk_+ \frac{f(k_-)\cdot k_+^n}{P_0\cdot P_1\cdots P_m} \;,\quad P_i=(k-p_i)_{||}^2-w_i+i\epsilon \;,
\end{align}
where $w_i$ and $p_i$ are constants with respect to $k_{\pm}$ and $f(k_-)$ is a function of $k_-$.

It has to be shown that by point-splitting, the integral
\begin{align}
\label{formps}
\int dk_- dk_+ \frac{f(k_-)\cdot k_+^n}{P_0\cdot P_1\cdots P_m}\;e^{i\xi^+ k_++i\xi^- k_-}\;,
\end{align}
for arbitrary $n,m$ gives the correctly evaluated result on the light-cone. The $k_+$-integration is always considered first in this section. 

In Section \ref{pointsplitting} we noted that the problems with calculating (\ref{D+ohne}) on the light-cone arise because of the hidden tadpole integrals which become visible after applying formula (\ref{id}). This identity is also applicable to a general integral (\ref{form}). It can be used successively until all powers of $k_+$ in the nominator have disappeared. For the case $n<m$ ($n>m$ cannot appear, for $n=m$ see below) this results in an integral that consists  of a sum of integrals of the type
\begin{align}
\label{afterid}
\int dk_- dk_+ \frac{F(k_-)}{P_0\cdot P_1\cdots P_i}\;,\quad 0\leq i\leq m\;.
\end{align}
 The dependence on $p_i$ and $w_i$ is suppressed in $F$. These integrals can be divided such that (\ref{form}) yields
\begin{align}
\int dk_- dk_+ \frac{f(k_-)\cdot k_+^n}{P_0\cdot P_1\cdots P_m}=\sum_i\int dk_- dk_+ \frac{F_i(k_-)}{ P_i}\;.
\end{align}
We know that the point-splitting procedure is capable of dealing with the integrals,
\begin{align}
\label{tadps}
\int dk_- dk_+ \frac{F_i(k_-)\,e^{i\xi^+ k_++i\xi^- k_-}}{P_i}\;,
\end{align}
since problems appear only because of the infrared light-cone singularities of $\frac{1}{k_-}$. After the insertion of the exponential in (\ref{tadps}) these (and also the usual ultraviolet divergences) are regularized. Therefore I conclude that a general integral of the form (\ref{form}) yields for $n<m$  the correct result after $k_+$- and $k_-$-integration when regularized with point-splitting.

In the previous arguments, it was implicitly assumed that all $p_i$ are different from each other. However, we also have to take care of integrals with s-fold tadpole integrands like  
\begin{align}
\label{s-fold}
\int dk_- dk_+ \frac{1}{ P_i^s}\quad,
\end{align}
since these integrals would disappear as well using contour integration without point-splitting. This is again due to the pole structure. Note that for $s>1$ these integrals are superficially finite, i.e. they are not ultraviolet divergent. In standard quantization there would be no regularization required in this case. 
Also these integrals are treated correctly by the point-splitting procedure. (\ref{s-fold}) can be recast by appropriate substitutions to yield
\begin{align}
I=e^{i\xi p_{i}}\int \frac{dk_-}{(2k_-)^s}\int dk_+ \frac{e^{i\xi^+ k_++i\xi^- k_-}}{(k_+-\frac{m^2-i\epsilon}{2k_-})^s}\quad.
\end{align} 
This integral leads to a representation of the modified Bessel function $K_{s-1}$,
\begin{align}
I&=-e^{i\xi p_{i}}\frac{2i\pi}{(s-1)!} \left({\xi_L^2}{4m^2}\right)^{\frac{s-1}{2}} \cdot K_{s-1}\left(m\xi_L \right) 
\end{align}
Once again the singularity for $k_-\to0$ is suitably regularized by point-splitting.
Performing the limit $\xi^{\pm}\to0$ the integral yields finally  
\begin{align}
\lim_{\xi^{\pm}\to0} I=\frac{(-1)^s}{s-1}\cdot\frac{i\pi}{(m^2)^{s-1}}\;\quad \mbox{for}\;  s>1\quad.
\end{align}

In the case $m=n$ additionally integrals like 
\begin{align}
\int dk_-F(k_-) \int dk_+ \frac{k_+\,e^{i\xi^+ k_++i\xi^- k_-}}{P_i}\quad,
\end{align}
might appear.
These can be rewritten as derivative with respect to $\xi^+$ and can therefore be reduced to the cases already treated.
 
In summary, to deal with the light-cone divergences in perturbative QED, I propose to introduce a point-splitting regularization for every internal momentum in any diagram (with different regulators). One has to consider first the integrals of the  subdiagrams. These are of the form (\ref{form}). After integration over each $k_+$ and performing the $k_-$-integrals that belong to tadpole contributions, one has to take the limit  $\xi^{\pm} \to 0$. The next integration is then also of the form (\ref{form}) etc. In the end this results in an expression where all $k_+$-integrations are carried out and which is equivalent to the result in standard coordinates.


\section{How to Implement Point-Splitting}

 An insertion of the proposed point-splitting procedure at a more general level is of course preferable. A possibility would be a change in the Feynman rules. The rule ``integrate over each internal momentum $\int\frac{d^4k}{(2\pi)^4}$'', could be altered to ``integrate over each internal momentum, insert a point-splitting exponential for the ${\pm}$-components and take the limit of vanishing regulator straight after the ${\pm}$-integration of that momentum''. Thus we set
\begin{align}
\lim_{\xi_k^{\pm}\to 0}\int\frac{d^4k}{(2\pi)^4}\;e^{ik_+\xi_k^++ik_-\xi_k^-} 
\end{align}
for each internal momentum, with different values of the regulators for every integral. These additional rules would ensure a correct treatment of light-cone singularities. The modified Feynman rules can also be introduced in canonical light-cone perturbation theory \cite{brodsky1,chang2,brodsky73} as
\begin{align}
\lim_{\xi_k^{\pm}\to 0}\int\frac{d^2k_{\perp}dk_-}{(2\pi)^3}\;\Theta(k_-)\;e^{i\frac{k_{\perp}^2+m^2}{2k_-}\xi_k^++ik_-\xi_k^-}\,. 
\end{align}

Of course, this kind of implementation of the procedure is not fully satisfactory. The Feynman rules originate from a Hamiltonian and  therefore the Hamiltonian or Lagrangian of the theory are the most appropriate place to implement point-splitting and thereby define a theory which is free of the light-cone singularities. 

The fact that the point-split expressions in Section \ref{pointsplitting} follow from a point-split matrix element of the interaction, suggests to point-split the operators of the QED Lagrangian. To preserve gauge invariance we include a Schwinger line integral
\begin{align}
U\left(x-\frac{\xi}{2},x+\frac{\xi}{2}\right)=\exp\left\{ie\int\limits_{x-\frac{\xi}{2}}^{x+\frac{\xi}{2}} dz^{\mu} A_{\mu}(z) \right\} \quad,
\end{align}
which transforms covariantly under $U(1)$ gauge transformations.
The proposal for the point-split Lagrangian reads then
\begin{align}
\label{psL}
{\cal L}=e\bar{\Psi}\left(x-\frac{\xi}{2}\right)\gamma^{\alpha} D_{\alpha} \left[U\left(x-\frac{\xi}{2},x+\frac{\xi}{2}\right)\Psi\left(x+\frac{\xi}{2}\right) \right],
\end{align}
where $D_{\alpha}=\partial_{\alpha}+ieA_{\alpha}$ is the covariant derivative. We see the connection between the point-splitting procedure from the previous sections and the proposal  to define order parameters as vacuum expectation values of equal light-cone time limits  of Heisenberg operators \cite{lenz3,burklenz}. If we calculate diagrams from the Lagrangian (\ref{psL}), we evaluate matrix elements of infinitesimally split Heisenberg operators.

\begin{figure}[t]
 \begin{center}
  \epsfig{file=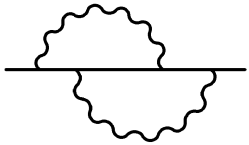, width=4cm, height = 2.4cm}
   \vspace{-0.4cm}
  \caption{Higher order diagram}
   \label{spoiler}
  \end{center}
\vspace{-0.4cm}\end{figure}

This implementation of point-splitting is however not fully equivalent to the proposed change in the Feynman rules since it does not lead to different regulators for each internal momentum. This yields the possibility of cancellations of point-splitting exponentials because of the $\delta$ functions enforcing momentum conservation at each vertex. This occurs for example in the diagram depicted in Fig.\ref{spoiler}. The part that lacks the exponential in this diagram consists however only of integrals of the form (\ref{form}) with $n<m-1$. The cancellations appear therefore only in those integrals that do not lead to additional problems on the light-cone even without point-splitting exponential. It still has to be investigated if this is true for all diagrams to any order. \\
\indent In \cite{osland} the interaction part of the Lagrangian in standard quantized QED is point-split using different fractions of a point-splitting parameter and therefore different regulators. Such an implementation would prevent cancellations of point-splitting exponentials and provides therefore another starting point for solving this problem.

\vspace{1cm}

\section{Summary}

In this work point-splitting was found to be a suitable procedure to obtain physically meaningful results in perturbative QED on the light-cone. It was shown that the insertion of a point-splitting exponential for light-cone energy and momentum in every integral avoids the light-cone problems in QED to any order.
The point-splitting procedure can be implemented by a change in the Feynman rules that leads to the insertion of different point-splitting regulators for every internal momentum. It would be desirable to implement point-splitting in a gauge invariant way in the Lagrangian of QED. Such an attempt has been proposed. Its general validity beyond $1$-loop order however  remains to be established.



\vspace{0.5cm}
\noindent {\bf Acknowledgments} 
I thank F. Lenz and M. Burkardt for many helpful and illuminating discussions. The author was supported by the Studienstiftung des deutschen Volkes.

 \end{document}